\documentclass[11pt,a4paper]{article}

\usepackage[utf8]{inputenc}
\usepackage[T1]{fontenc}
\usepackage{lmodern}
\usepackage[margin=2.5cm]{geometry}
\usepackage{graphicx}
\usepackage{booktabs}
\usepackage{multirow}
\usepackage{longtable}
\usepackage{xcolor}
\usepackage{hyperref}
\usepackage{amsmath}
\usepackage{enumitem}
\usepackage{caption}
\usepackage{subcaption}
\usepackage{float}
\usepackage{tabularx}
\usepackage{array}
\usepackage{microtype}
\usepackage{url}
\usepackage{authblk}
\usepackage{tikz}
\usetikzlibrary{shapes.geometric, arrows.meta, positioning, fit, backgrounds}

\hypersetup{
    colorlinks=true,
    linkcolor=blue!60!black,
    citecolor=blue!60!black,
    urlcolor=blue!60!black
}
\newcolumntype{R}{>{\raggedleft\arraybackslash}p{1.2cm}}

\title{\textbf{Software Testing at the Network Layer:}\\[0.3em]
\Large Automated HTTP API Quality Assessment and Security\\
Analysis of Production Web Applications}

\author[1,2]{Ali Hassaan Mughal}
\author[1,3]{Muhammad Bilal}
\author[1,4]{Noor Fatima}
\affil[1]{Independent Researcher}
\affil[2]{Applied-MBA \ Data Analytics, Texas Wesleyan University, USA}
\affil[3]{M.Sc.\ Management, Technical University of Munich, Germany}
\affil[4]{B.E. \ Computer Engineering, National University of Sciences and Technology, Pakistan}
\date{February 2026}

\begin{document}
\maketitle

\begin{abstract}
Modern web applications rely heavily on client-side API calls to fetch data, render content, and communicate with backend services. However, the quality of these network interactions (redundant requests, missing cache headers, oversized payloads, and excessive third-party dependencies) is rarely tested in a systematic way. Moreover, many of these quality deficiencies carry security implications: missing cache headers enable cache poisoning, excessive third-party dependencies expand the supply-chain attack surface, and error responses risk leaking server internals. In this study, we present an automated software testing framework that captures and analyzes the complete HTTP traffic of \textbf{18 production websites} spanning 11 categories (e-commerce, news, government, developer tools, travel, and more). Using automated browser instrumentation via Playwright, we record 108 HAR (HTTP Archive) files across 3 independent runs per page, then apply 8 heuristic-based anti-pattern detectors to produce a composite quality score (0--100) for each site. Our results reveal a wide quality spectrum: minimalist server-rendered sites achieve perfect scores of 100, while content-heavy commercial sites score as low as 56.8. We identify redundant API calls and missing cache headers as the two most pervasive anti-patterns, each affecting 67\% of sites, while third-party overhead exceeds 20\% on 72\% of sites. One utility site makes 2,684 requests per page load, which is 447$\times$ more than the most minimal site. To protect site reputations, all identities are anonymized using category-based pseudonyms. We provide all analysis scripts, anonymized results, and reproducibility instructions as an open artifact. This work establishes an empirical baseline for HTTP API call quality across the modern web and offers a reproducible testing framework that researchers and practitioners can apply to their own applications.
\end{abstract}

\textbf{Keywords:} software testing, HTTP API quality, anti-pattern detection, web application security, network traffic analysis, HAR analysis, automated testing, web quality scoring, third-party dependencies, empirical software engineering

\section{Introduction}
\label{sec:introduction}

Every time a user visits a modern website, their browser makes dozens, sometimes hundreds, of HTTP requests behind the scenes. Some fetch the page's HTML; others load images, stylesheets, and JavaScript bundles. A growing share, however, are \emph{API calls}: XHR and Fetch requests that retrieve JSON data from backend services, third-party analytics platforms, advertising networks, and social media integrations.

The quality of these API calls matters for both performance and security. Redundant requests waste bandwidth and server resources. Missing cache headers force browsers to re-fetch data that has not changed, while also leaving responses vulnerable to cache poisoning by intermediary proxies. Oversized JSON payloads transfer far more data than the client requires, potentially exposing sensitive fields to network-level attackers. Excessive third-party calls create privacy concerns, single points of failure~\cite{kashaf2020analyzing}, and an expanded supply-chain attack surface. Despite their importance, however, API call patterns in production websites are rarely tested or audited in a systematic way.

Traditional software testing approaches focus on functional correctness at the UI layer~\cite{mughal2025autonomous, li2024survey, kertusha2025survey} or on API design compliance at the specification level~\cite{palma2014detection, rodriguez2016rest, bogner2023rest, kim2022restapi, golmohammadi2023restsurvey}. Web performance research has examined page load times~\cite{wang2013demystifying}, resource prioritization~\cite{netravali2016polaris}, and third-party tracking~\cite{englehardt2016census}. However, these studies address either \emph{what} is loaded or \emph{how APIs are designed}, rather than \emph{how efficiently the loading is orchestrated at runtime} and what quality and security implications arise from the observed patterns. We address this gap by posing two research questions: \textbf{(RQ1)} How good are the HTTP API call patterns of popular production websites? \textbf{(RQ2)} What are the quality and security implications of the anti-patterns observed?

To answer this, we designed and executed a three-phase automated testing study:

\begin{enumerate}[leftmargin=2em]
    \item \textbf{Capture}: We used Playwright, a browser automation framework~\cite{garcia2024exploring}, to visit 18 production websites and record their complete HTTP traffic as HAR (HTTP Archive) files, performing 3 independent runs per page across 2 pages per site to yield 108 capture files.
    \item \textbf{Analyze}: We developed 8 heuristic-based anti-pattern detectors that examine each HAR file for common API call quality issues: redundant calls, N+1 query patterns, missing cache headers, oversized payloads, missing compression, excessive third-party overhead, sequential waterfalls, and error responses.
    \item \textbf{Score}: We combined the 8 anti-pattern dimensions into a weighted composite quality score (0--100), where 100 indicates no detected anti-patterns.
\end{enumerate}

Our contributions are:
\begin{itemize}[leftmargin=2em]
    \item An \textbf{empirical dataset} of 108 HAR captures from 18 production websites across 11 categories.
    \item A \textbf{reproducible automated testing framework} with 8 anti-pattern detectors and a composite scoring rubric.
    \item \textbf{Quantitative findings} showing a wide quality spectrum (56.8--100), with redundant calls and missing cache headers as the most pervasive issues.
    \item A \textbf{security analysis} of the detected anti-patterns, linking quality deficiencies to concrete security risks including supply-chain vulnerabilities and information leakage.
    \item A \textbf{public artifact} (code, anonymized data, and instructions) enabling replication and extension.
\end{itemize}

\section{Related Work}
\label{sec:related}

Our work spans several research areas: web performance measurement, third-party resource analysis, API design quality, network traffic analysis, frontend architecture, software quality and security assessment, and empirical methodology. We review each in turn.

\subsection{Web Performance Measurement}

The HTTP Archive project~\cite{httparchive} has tracked resource loading trends across millions of websites since 2010, providing aggregate statistics on page weight, request counts, and technology adoption. Butkiewicz et al.~\cite{butkiewicz2011characterizing} characterized web page complexity and its impact on load times, establishing that page complexity, rather than network latency, is the dominant factor in modern page loads. Wang et al.~\cite{wang2013demystifying} developed WProf to demystify the dependency structure of page loads, revealing that computation and object dependencies account for 35\% of critical-path time. Netravali et al.~\cite{netravali2016polaris} exploited fine-grained dependency tracking with Polaris to accelerate page loads by 34\% on average.

Ihm and Pai~\cite{ihm2011towards} provided an early longitudinal analysis of web traffic, revealing a 55\% reduction in content cachability between 2006 and 2010, driven by increasing personalization and dynamic content. Bocchi et al.~\cite{bocchi2016measuring} proposed automated approaches for measuring the Quality of Experience (QoE) of web users, establishing metrics beyond raw load times. Sundaresan et al.~\cite{sundaresan2013measuring} demonstrated that network infrastructure and ISP configurations significantly affect web performance in broadband access networks. Pourghassemi et al.~\cite{pourghassemi2019what} applied causal profiling to web browsers, enabling ``what-if'' analysis that identifies which browser subsystems are performance bottlenecks. Aqeel et al.~\cite{aqeel2020landing} revealed the ``Jekyll and Hyde'' phenomenon where landing pages differ drastically in performance from internal pages, a finding directly relevant to our two-page-per-site capture design.

More recently, Google's Core Web Vitals~\cite{webvitals2020} have established industry-standard performance metrics (Largest Contentful Paint, First Input Delay, Cumulative Layout Shift) that focus on user-perceived quality. Netravali et al.~\cite{netravali2018vesper} proposed Vesper to measure time-to-interactivity as a new quality dimension beyond simple load times. Our work complements these metrics by examining the \emph{API call patterns} underlying the performance numbers, measuring not just how fast a page loads but how efficiently its network requests are orchestrated.

\subsection{Third-Party Resource Analysis}

Third-party resources represent a significant and growing share of web traffic. Englehardt and Narayanan~\cite{englehardt2016census} conducted a census of one million websites, identifying tracking behaviors across 81,000 third-party domains. Nikiforakis et al.~\cite{nikiforakis2012you} evaluated remote JavaScript inclusions at scale, finding that 88.5\% of the Alexa Top 10,000 include at least one external script, creating transitive trust dependencies. Mayer and Mitchell~\cite{mayer2012third} provided an early comprehensive policy and technology analysis of third-party web tracking.

Lerner et al.~\cite{lerner2016internet} conducted an ``archaeological'' study of web tracking from 1996 to 2016, documenting the evolution of tracking technologies from simple cookies to sophisticated fingerprinting. Bashir and Wilson~\cite{bashir2018diffusion} traced how user tracking data diffuses through the advertising ecosystem, revealing complex chains of data sharing. Kashaf et al.~\cite{kashaf2020analyzing} analyzed third-party service dependencies in the wake of the Mirai-Dyn DDoS incident, finding that top websites depend on a small number of critical infrastructure providers, making third-party overhead not merely a performance issue but a resilience and security risk. Ikram et al.~\cite{ikram2019trust} further demonstrated that approximately 50\% of first-party websites render content they did not directly load, revealing deep transitive trust chains that amplify the supply-chain risk. At the detection level, Iqbal et al.~\cite{iqbal2020adgraph} demonstrated that graph-based machine learning can classify ad and tracker resources with 95\% accuracy.

Our third-party overhead detector builds on similar domain classification techniques but extends the analysis to measure the \emph{proportion} of third-party requests and categorize them (analytics, ads, social, CDN, tracking). Unlike prior work that focuses on privacy or tracking, we assess third-party requests as an \emph{API quality dimension}.

\subsection{HTTP Caching, Compression, and Protocol Evolution}

Effective caching and compression are fundamental to API quality. Vesuna et al.~\cite{vesuna2016caching} investigated mobile web caching effectiveness, finding that caching alone provides modest speedups due to the dominance of dynamic content. Agababov et al.~\cite{agababov2015flywheel} presented Google's Flywheel data compression proxy for mobile web, achieving 50\% byte savings through server-side transcoding and compression.

CDN infrastructure plays a critical role in API delivery. Nygren et al.~\cite{nygren2010akamai} described the Akamai CDN platform's architecture for high-performance content delivery. Calder et al.~\cite{calder2015analyzing} analyzed anycast CDN performance, revealing that DNS-based redirection sometimes routes clients to suboptimal servers.

Protocol-level improvements also affect API quality. Wijnants et al.~\cite{wijnants2018http2} studied HTTP/2 prioritization and its impact on web performance, demonstrating that improper priority assignment can negate multiplexing benefits. Marx et al.~\cite{marx2020resource} compared resource multiplexing and prioritization between HTTP/2 over TCP and HTTP/3 over QUIC, finding that QUIC's independent streams mitigate head-of-line blocking.

Our missing cache headers and missing compression detectors directly measure whether sites leverage these well-understood optimization techniques.

\subsection{REST API Design and Anti-Patterns}

At the design level, Palma et al.~\cite{palma2014detection} proposed SODA-R for automated detection of REST API (anti)patterns, achieving over 75\% precision and recall across 5 patterns and 8 anti-patterns. Rodriguez et al.~\cite{rodriguez2016rest} analyzed 78\,GB of HTTP traffic to assess real-world REST API compliance with design best practices. Petrillo et al.~\cite{petrillo2016api} compiled a catalog of 73 REST API design best practices and assessed cloud APIs against them.

More recently, Bogner et al.~\cite{bogner2023rest} conducted a controlled experiment with 105 participants demonstrating that violations of 11 of 12 REST design rules significantly worsened API comprehension. Tran et al.~\cite{tran2021formalising} formalized REST API practices as design (anti)patterns using the SARA approach. Bogner et al.~\cite{bogner2024microservice} extended this to microservice API patterns, finding measurable impacts on understandability.

The REST vs.\ GraphQL debate is also relevant: Brito and Valente~\cite{brito2020rest} found that GraphQL requires less development effort than REST, while Seabra et al.~\cite{seabra2019rest} showed GraphQL improved performance in two-thirds of tested applications. API evolution studies~\cite{xavier2017historical, hora2018developers, lercher2024microservice} document how APIs change over time and the challenges this creates for consumers.

At the testing level, Kim et al.~\cite{kim2022restapi} compared 10 automated REST API testing tools at ISSTA, finding significant gaps in test coverage and fault detection; Golmohammadi et al.~\cite{golmohammadi2023restsurvey} surveyed 92 studies on REST API testing in TOSEM, yet none address the runtime quality of API call orchestration patterns. On the server side, Hubener et al.~\cite{hubener2022microservice} detect similar anti-patterns (e.g., chatty services, N+1 patterns) in microservice architectures using distributed tracing at ICSE.

Our work complements this design-level and testing-level analysis by measuring the \emph{runtime manifestation} of API anti-patterns in production traffic from the client side, bridging the gap between API design theory and actual network behavior.

\subsection{JavaScript and Frontend Architecture}

The choice of frontend architecture significantly impacts API call patterns. Selakovic and Pradel~\cite{selakovic2016performance} conducted an empirical study of JavaScript performance issues, identifying common anti-patterns in production code. Pano et al.~\cite{pano2018factors} studied factors leading to JavaScript framework adoption, documenting the performance trade-offs between different framework choices.

Netravali and Mickens~\cite{netravali2018prophecy} proposed Prophecy, which uses final-state write logs to accelerate mobile page loads, highlighting the overhead of client-side JavaScript execution. Mardani et al.~\cite{mardani2020fawkes} demonstrated 46--64\% faster page loads through static templating that separates layout from dynamic API content (NSDI 2020), while Ko et al.~\cite{ko2021oblique} used symbolic execution to identify unnecessary JavaScript computation (NSDI 2021). Mardani et al.~\cite{mardani2021horcrux} quantified the JavaScript serial execution bottleneck on modern pages (OSDI 2021), providing the systems-level explanation for performance degradation in JavaScript-heavy architectures. Iskandar et al.~\cite{iskandar2020comparison} compared client-side and server-side rendering approaches, finding that SSR provides faster initial page loads but shifts computational cost to servers. These architectural decisions directly influence the volume and pattern of API calls we observe in our study.

\subsection{Network Traffic Analysis and HAR-Based Studies}

HAR (HTTP Archive) files provide a standardized format for recording HTTP traffic. Bhole and Popescu~\cite{bhole2005measurement} provided foundational HTTP traffic characterization covering request sizes, response codes, and transaction counts. Bermbach and Wittern~\cite{bermbach2016benchmarking} proposed a benchmark framework for measuring web API quality including performance, availability, and compliance; their extended longitudinal study~\cite{bermbach2020revisited} added security assessment dimensions and tracked API quality changes over three years. Xavier~\cite{xavier2024web} conducted a large-scale analysis of 1.3 million domains, finding that approximately 50\% of all web traffic is concentrated in just 3,000 websites.

Several commercial tools (Lighthouse, WebPageTest, GTmetrix) perform single-page audits but do not systematically compare API patterns across a curated site corpus. Our work fills this gap with a multi-site, multi-run analysis using a consistent and reproducible methodology.

\subsection{Software Quality, Testing, and Anti-Pattern Detection}

Software quality research provides methodological foundations for our work. Sharma and Spinellis~\cite{sharma2018survey} surveyed 445 primary studies on software smells, identifying five defining characteristics: indicator of poor solution, violation of best practices, quality impact, and recurrence. Garousi and Kucuk~\cite{garousi2018smells} produced the largest catalogue of test smells from 166 sources.

Web application testing has advanced significantly in recent years. Li et al.~\cite{li2024survey} surveyed 314 primary studies on web application testing spanning 2014--2023, documenting evolution in testing techniques. Kertusha et al.~\cite{kertusha2025survey} analyzed 214 papers on web testing, identifying Selenium as the most widely used tool and noting limited industrial adoption of academic techniques. Mughal~\cite{mughal2025autonomous} proposed an autonomous reinforcement learning agent for dynamic web UI testing within a Behavior-Driven Development framework, demonstrating improvements in defect detection and test coverage. Kim et al.~\cite{kim2023nlprest} further advanced automated API analysis by applying NLP to specification-based test generation (ISSTA 2023). Our work extends automated quality assessment from the UI layer to the network API layer, and the two approaches are complementary: UI testing validates functional behavior while API quality auditing validates network efficiency.

\subsection{Web Measurement Methodology}

Rigorous empirical methodology is essential for web measurement studies. Wohlin et al.~\cite{wohlin2012experimentation} provide the standard reference for empirical study design in software engineering. Runeson and H{\"o}st~\cite{runeson2009guidelines} established guidelines for case study research in software engineering. Kitchenham and Charters~\cite{kitchenham2007guidelines} defined the methodology for systematic literature reviews in the field.

Particularly relevant to our work, Demir et al.~\cite{demir2022reproducibility} surveyed 117 web measurement papers and found that 73\% omit which pages were analyzed and only 41.9\% of reproducibility criteria were satisfied, highlighting the importance of methodological transparency. In a follow-up study, Demir et al.~\cite{demir2023similarity} visited 1.7 million pages with five different setups (IMC 2023), finding that even identical configurations produce divergent results, reinforcing the need for multi-run protocols. Jueckstock et al.~\cite{jueckstock2021crawl} systematically investigated how crawling setup choices (browser type, automation tool, network position) affect web measurement results (WWW 2021), finding significant differences even between identical setups. We address these concerns by documenting our exact capture configuration, page selection, and analysis logic, and by releasing our complete analysis pipeline. Garcia et al.~\cite{garcia2024exploring} compared four major browser automation frameworks (Selenium, Cypress, Puppeteer, Playwright), informing our choice of Playwright for its superior HAR recording capabilities.

Goel et al.~\cite{goel2022jawa} demonstrated that JavaScript-heavy pages are inherently difficult to capture faithfully due to non-deterministic execution (OSDI 2022), validating the need for multi-run capture protocols. Their follow-up work, Sprinter~\cite{goel2024sprinter}, characterized JavaScript-induced non-determinism in web crawls at scale (NSDI 2024). Olston and Najork~\cite{olston2010web} provided a comprehensive survey of web crawling science, while Stafeev and Pellegrino~\cite{stafeev2024sok} evaluated crawling algorithms for web security measurements, finding that randomized approaches can surprisingly outperform designed algorithms. Chiapponi~\cite{chiapponi2023detecting} studied bot detection techniques relevant to our automated browsing methodology. Shull et al.~\cite{shull2008role} discussed the role of replications in building reliable empirical knowledge, supporting our three-run-per-page design.

\section{Methodology}
\label{sec:methodology}

\begin{figure}[H]
\centering
\includegraphics[width=\textwidth]{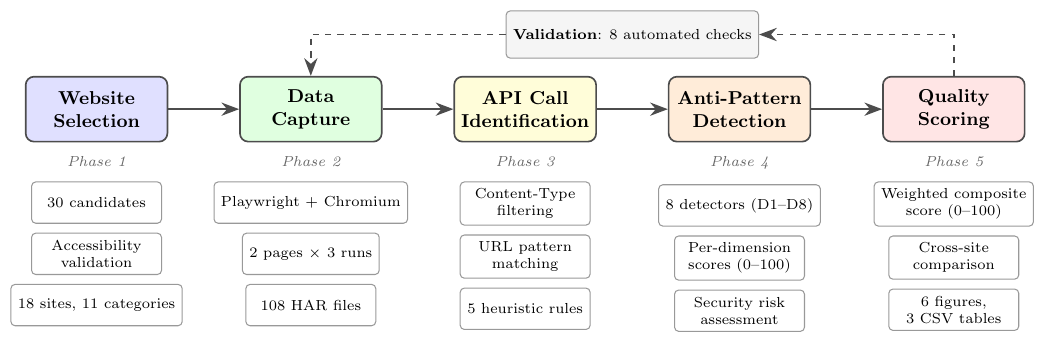}
\caption{Five phases of evaluation for the API quality and accumulated calculation of quality score through a data capture to quality scoring loop.}
\label{fig:methodology}
\end{figure}

This section describes how we selected websites, captured their network traffic, detected anti-patterns, and computed quality scores. Every step is fully automated and reproducible. Figure~\ref{fig:methodology} provides an overview of the complete pipeline.

\subsection{Website Selection}
\label{sec:site-selection}

We selected 30 candidate websites using the following criteria:

\begin{enumerate}[leftmargin=2em]
    \item \textbf{Diversity of category}: We targeted diverse categories to ensure breadth: e-commerce, news, government, developer tools, travel, reference, forum, classifieds, utility, entertainment, and developer blog.
    \item \textbf{Diversity of architecture}: We deliberately included sites with different technology stacks, ranging from minimal server-rendered HTML and progressive enhancement to React SSR with hydration, Next.js SSR, and heavy client-side rendering.
    \item \textbf{Popularity}: All selected sites rank in the Similarweb Top 500 for their respective categories.
    \item \textbf{Public accessibility}: We excluded sites requiring authentication (login walls) or paid subscriptions.
\end{enumerate}

\textbf{Anonymization.} To protect the reputations of the tested websites, we anonymize all site identities using category-based pseudonyms throughout this paper and in all published data. This is standard practice in empirical software engineering research when studies assign evaluative scores~\cite{runeson2009guidelines}. Our quality scores reflect a single measurement snapshot from one geographic location on one date; they should not be interpreted as definitive judgments on the engineering quality of any organization. Sites may have valid architectural reasons for observed patterns (e.g., ad-revenue requirements, A/B testing, personalization engines), and web applications change frequently.

\textbf{Accessibility validation.} Before data collection, we ran an automated validation pass that visited each site's homepage and classified the result as \texttt{success}, \texttt{blocked}, or \texttt{timeout}. Sites returning HTTP 401/403/429 or displaying CAPTCHA/bot-detection pages were classified as blocked. Sites that did not reach Playwright's \texttt{networkidle} state within 90 seconds were classified as timeout.

Of the 30 candidates:
\begin{itemize}[leftmargin=2em]
    \item \textbf{19 accessible}: Successfully loaded with HTTP 200 and meaningful content.
    \item \textbf{5 blocked}: Two social media sites, one marketplace, one review site, and one real-estate site returned HTTP 403.
    \item \textbf{6 timed out}: Five sites with heavy SPA architectures and one public news site exceeded the 90-second timeout.
\end{itemize}

One additional site returned HTTP 200 but with only 6 requests and a paywall redirect (HTTP 401 for all content requests), leaving effectively no analyzable traffic. We excluded this site during analysis, yielding a final corpus of \textbf{18 sites} across 11 categories. Table~\ref{tab:site-corpus} lists the complete corpus with architectural annotations.

\begin{table}[t]
\centering
\caption{Website corpus: 18 production sites across 11 categories, ordered by quality score. All names are category-based pseudonyms.}
\label{tab:site-corpus}
\small
\begin{tabularx}{\textwidth}{@{}l l X r@{}}
\toprule
\textbf{Site} & \textbf{Category} & \textbf{Architecture} & \textbf{Score} \\
\midrule
Forum-1        & Forum         & Minimal server-rendered          & 100.0 \\
Government-1   & Government    & Server-rendered, progressive enh.& 100.0 \\
Classifieds-1  & Classifieds   & Minimal server-rendered HTML     & 97.5 \\
DevBlog-1      & Dev Blog      & Rails + Preact (Forem)           & 95.4 \\
Reference-1    & Reference     & Server-rendered, minimal JS      & 92.2 \\
DevTool-1      & Dev Tools     & Server-rendered + Turbo/Stimulus & 85.4 \\
Reference-2    & Reference     & Server-rendered, legacy          & 84.3 \\
Commerce-2     & E-commerce    & Server-rendered + Marko.js       & 81.4 \\
DevTool-2      & Dev Tools     & Server-rendered + jQuery         & 78.7 \\
News-3         & News          & React SSR + heavy JS             & 70.9 \\
News-2         & News          & Heavy client-side rendering      & 69.6 \\
Commerce-1     & E-commerce    & Server-rendered + heavy JS       & 65.3 \\
Entertainment-1& Entertainment & Next.js SSR                      & 63.6 \\
Travel-2       & Travel        & React SSR + client hydration     & 61.8 \\
Travel-1       & Travel        & Server-rendered + heavy JS       & 61.1 \\
Commerce-3     & E-commerce    & React SPA + SSR                  & 60.8 \\
Utility-1      & Utility       & Heavy ads + API-driven           & 59.9 \\
News-1         & News          & Server-rendered + hydration       & 56.8 \\
\bottomrule
\end{tabularx}
\end{table}

\subsection{Data Collection}
\label{sec:data-collection}

We used Playwright (v1.49, Chromium engine)~\cite{garcia2024exploring} to capture HAR files with the following configuration:

\begin{itemize}[leftmargin=2em]
    \item \textbf{Browser}: Chromium 131, headless mode, with automation detection disabled \\ (\texttt{--disable-blink-features=AutomationControlled}).
    \item \textbf{Viewport}: 1920$\times$1080 pixels (standard desktop).
    \item \textbf{User agent}: Chrome 131 on Windows 10 (standard desktop string).
    \item \textbf{Pages per site}: 2 (typically homepage + a content/search page).
    \item \textbf{Runs per page}: 3 independent visits, each using a fresh browser context (no cached data or cookies), following replication guidelines~\cite{shull2008role}.
    \item \textbf{Visit protocol}: For each visit, we: (1) navigate to the URL and wait for \texttt{DOMContentLoaded}, (2) wait up to 5\,s for network idle, (3) scroll 3 viewport heights to trigger lazy loading, (4) wait 3\,s for lazy-loaded API calls, then (5) scroll back to top.
    \item \textbf{HAR content}: Headers only; response bodies are omitted to reduce storage and avoid capturing user content.
    \item \textbf{Timeout}: A hard timeout of 90 seconds per page prevents infinite SPA loading~\cite{chiapponi2023detecting}. If triggered, the browser context is force-closed.
    \item \textbf{Inter-run delay}: 2 seconds between runs, 3 seconds between sites.
\end{itemize}

This protocol produced \textbf{108 HAR files} (18 sites $\times$ 2 pages $\times$ 3 runs) totaling approximately 175\,MB of header data. All captures were performed from a single residential IP address in the United States on February 7, 2026.

\subsection{API Call Identification}
\label{sec:api-identification}

Not every HTTP request constitutes an API call. Static assets such as images, CSS stylesheets, and fonts are resource loads rather than API interactions. To separate API calls from static resource requests, we classify a request as an API call if \emph{any} of the following five heuristics match:

\begin{enumerate}[leftmargin=2em]
    \item The response \texttt{Content-Type} header contains \texttt{application/json} or \texttt{application/graphql}.
    \item The request includes \texttt{X-Requested-With: XMLHttpRequest}.
    \item The request \texttt{Accept} header contains \texttt{application/json}.
    \item The request \texttt{Content-Type} header contains \texttt{application/json} (POST/PUT payloads).
    \item The URL path matches known API patterns: \texttt{/api/}, \texttt{/graphql}, \texttt{/v1/}, \texttt{/v2/}, \texttt{/v3/}, \texttt{/ajax/}, \texttt{/rest/}, \texttt{/\_next/data/}, or \texttt{/wp-json/}.
\end{enumerate}

This heuristic approach trades some precision for practical recall, as it captures the vast majority of genuine API calls while accepting occasional false positives from non-standard Content-Type usage. Similar heuristic classification approaches have been used in prior HTTP traffic characterization studies~\cite{bhole2005measurement, rodriguez2016rest}.

\subsection{Anti-Pattern Detectors}
\label{sec:detectors}

We implemented 8 detectors (D1--D8), each targeting a specific class of API call quality issue. Our detector design draws on REST anti-pattern catalogs~\cite{palma2014detection, bogner2023rest} and software smell frameworks~\cite{sharma2018survey}. For each detector, we describe the detection logic, the severity metric, and the scoring formula that produces a 0--100 dimension score. A score of 100 indicates no instances of the anti-pattern, while lower scores indicate increasing severity.

\subsubsection{D1: Redundant API Calls}
\textbf{What it detects}: Identical API calls (same HTTP method + URL) made multiple times within a single page load.\\
\textbf{Logic}: Count the number of unique \texttt{METHOD:URL} pairs called 2+ times among API requests. The \emph{excess} is the total number of duplicate calls minus 1 per unique URL.\\
\textbf{Scoring}: $\text{score} = \max(0,\ 100 - \text{excess} \times 10)$. Ten or more excess calls yield a score of 0.

\subsubsection{D2: N+1 Query Patterns}
\textbf{What it detects}: Bursts of API calls to the same URL pattern with varying path parameters (e.g., \texttt{/api/users/1}, \texttt{/api/users/2}, ..., \texttt{/api/users/N}).\\
\textbf{Logic}: Replace numeric path segments with \texttt{\{id\}}, group by pattern, and flag patterns with 3+ distinct URLs.\\
\textbf{Scoring}: $\text{score} = \max(0,\ 100 - \text{patterns} \times 20)$. Five or more N+1 patterns yield 0.

\subsubsection{D3: Sequential Waterfalls}
\textbf{What it detects}: API calls to the same domain that could potentially have been parallelized but were executed sequentially.\\
\textbf{Logic}: Sort API calls by start time, identify consecutive same-domain calls to different endpoints, and estimate wasted time.\\
\textbf{Scoring}: $\text{score} = \max(0,\ 100 - \text{wasted\_ms} / 50)$.

\subsubsection{D4: Missing Cache Headers}
\textbf{What it detects}: API responses (HTTP 200) that provide no caching guidance, specifically no \texttt{Cache-Control}, no \texttt{ETag}, and no \texttt{Last-Modified} header.\\
\textbf{Logic}: Count the percentage of successful API responses missing all three cache-related headers.\\
\textbf{Scoring}: $\text{score} = \max(0,\ 100 - \text{missing\_\%})$. If 100\% of API responses lack cache headers, the score is 0. This detector is motivated by findings of declining cache effectiveness~\cite{ihm2011towards, vesuna2016caching}.

\subsubsection{D5: Oversized Payloads}
\textbf{What it detects}: API responses larger than 100\,KB, suggesting over-fetching or missing pagination.\\
\textbf{Logic}: Count API responses with \texttt{bodySize} > 100,000 bytes.\\
\textbf{Scoring}: $\text{score} = \max(0,\ 100 - \text{count} \times 15)$.

\subsubsection{D6: Missing Compression}
\textbf{What it detects}: API responses larger than 1\,KB without \texttt{Content-Encoding} (gzip, br, deflate).\\
\textbf{Logic}: Estimate potential savings at 70\% compression ratio.\\
\textbf{Scoring}: $\text{score} = \max(0,\ 100 - \text{savings\_KB} / 5)$. This detector builds on compression effectiveness studies~\cite{agababov2015flywheel}.

\subsubsection{D7: Third-Party Overhead}
\textbf{What it detects}: The proportion of requests going to third-party domains (analytics, ads, social widgets, CDNs, trackers).\\
\textbf{Logic}: Extract the registered domain from each request's host and classify as first-party (matches the site's main domain) or third-party. Third-party requests are further categorized into analytics, ads, social, CDN, tracking, or other using a keyword dictionary of 40+ known third-party domain patterns~\cite{englehardt2016census, nikiforakis2012you}.\\
\textbf{Scoring}: $\text{score} = \max(0,\ 100 - \text{third\_party\_\%})$. A site where 100\% of requests are third-party scores 0.

\subsubsection{D8: Error Responses}
\textbf{What it detects}: API calls returning HTTP 4xx or 5xx status codes.\\
\textbf{Logic}: Count the percentage of API calls with status $\geq$ 400.\\
\textbf{Scoring}: $\text{score} = \max(0,\ 100 - \text{error\_\%} \times 5)$. A 20\% error rate yields 0.

\subsection{Composite Quality Score}
\label{sec:scoring}

The composite quality score is a weighted average of the 8 dimension scores. Weights were assigned based on practical impact and prevalence in web performance literature~\cite{httparchive, sharma2018survey}:

\begin{table}[H]
\centering
\caption{Quality score dimension weights.}
\label{tab:weights}
\begin{tabular}{@{}lr@{}}
\toprule
\textbf{Dimension} & \textbf{Weight (\%)} \\
\midrule
D1: Redundant Calls         & 15 \\
D2: N+1 Patterns            & 10 \\
D3: Sequential Waterfalls   & 10 \\
D4: Missing Cache Headers   & 15 \\
D5: Oversized Payloads      & 15 \\
D6: Missing Compression     & 10 \\
D7: Third-Party Overhead    & 15 \\
D8: Error Rate              & 10 \\
\midrule
\textbf{Total}              & \textbf{100} \\
\bottomrule
\end{tabular}
\end{table}

The composite score for each HAR capture is:
\[
Q = \sum_{i=1}^{8} w_i \cdot \min(100,\ \max(0,\ S_i))
\]
where $w_i$ is the weight of dimension $i$ (summing to 1.0) and $S_i$ is the dimension score. Per-site scores are averaged across all captures (3 runs $\times$ 2 pages = 6 captures per site).

\subsection{Data Validation}
\label{sec:validation}

To ensure correctness of the published results, we implemented an independent validation script (\texttt{validate.py}) that performs 8 automated checks, following empirical validation best practices~\cite{wohlin2012experimentation}:

\begin{enumerate}[leftmargin=2em]
    \item \textbf{Request count verification}: Re-parses all HAR files independently and compares raw entry counts against reported averages.
    \item \textbf{Invalid site detection}: Flags sites where all captures returned non-200 HTTP statuses (e.g., HTTP 401 paywall redirects).
    \item \textbf{Scoring formula verification}: Recalculates composite scores from dimension scores and checks for deviations > 0.2 points.
    \item \textbf{Domain extraction audit}: Verifies that multi-part TLDs (e.g., \texttt{.co.uk}) are handled correctly.
    \item \textbf{CSV--JSON consistency}: Confirms that CSV tables match per-site JSON reports.
    \item \textbf{Score sanity checks}: Verifies that minimal sites score high and heavy sites score lower.
    \item \textbf{HAR file completeness}: Ensures all 18 sites have the expected 6 HAR files each (2 pages $\times$ 3 runs).
    \item \textbf{Run-to-run consistency}: Checks that request counts across independent runs follow expected patterns.
\end{enumerate}

All 8 checks passed on the final dataset, with the only notes being expected Jensen's inequality effects in averaged non-linear scores (a known property of non-linear aggregation functions).

\section{Results}
\label{sec:results}

\subsection{Overall Quality Distribution}
\label{sec:overall}

\begin{figure}[t]
    \centering
    \includegraphics[width=\textwidth]{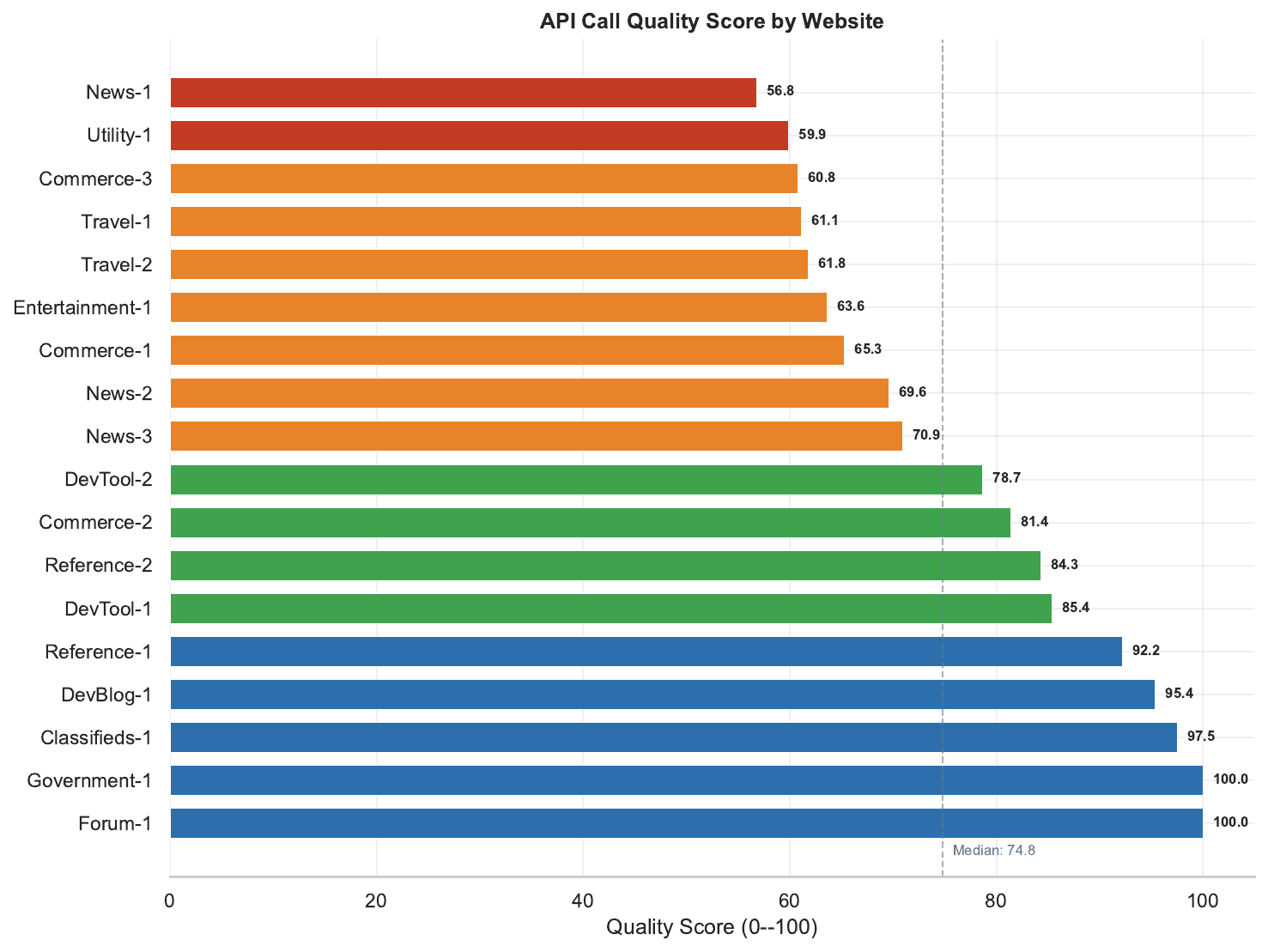}
    \caption{API call quality scores for 18 production websites. Colors indicate score tiers: red ($<$60), orange (60--75), green (75--90), blue (90+). The dashed line marks the median (74.8).}
    \label{fig:quality-scores}
\end{figure}

Figure~\ref{fig:quality-scores} presents the quality scores for all 18 sites, sorted from lowest to highest. The scores range from \textbf{56.8} (News-1) to \textbf{100.0} (Government-1, Forum-1), with a mean of \textbf{76.9} (median 74.8, $\sigma$ = 15.4).

The distribution reveals a clear architectural divide:
\begin{itemize}[leftmargin=2em]
    \item \textbf{Top tier (90--100):} Minimal or server-rendered sites with few or no API calls: Forum-1 (6 requests, score 100), Government-1 (13.5 requests, score 100), Classifieds-1 (68.5 requests, score 97.5), DevBlog-1 (62.7 requests, score 95.4), and Reference-1 (59.5 requests, score 92.2).
    \item \textbf{Middle tier (75--90):} Traditional server-rendered sites with moderate JavaScript: DevTool-1 (138.5 requests, score 85.4), Reference-2 (166.5 requests, score 84.3), Commerce-2 (231.7 requests, score 81.4), DevTool-2 (349.3 requests, score 78.7).
    \item \textbf{Lower tier ($<$75):} Heavy JavaScript SPAs and ad-driven sites: News-1 (681.8 requests, score 56.8), Utility-1 (2,683.7 requests, score 59.9), Commerce-3 (433.2 requests, score 60.8), Travel-1 (297.3 requests, score 61.1), Travel-2 (248.3 requests, score 61.8).
\end{itemize}

\subsection{Request Volume and Page Size}
\label{sec:volume}

Table~\ref{tab:full-results} reports the full results for all 18 sites. Request counts vary by two orders of magnitude: Forum-1 averages just \textbf{6 requests} per page load, while Utility-1 averages \textbf{2,683.7}, a 447$\times$ difference. Page sizes show similar disparity: Forum-1 transfers 12.7\,KB while News-3 transfers 25,816.5\,KB (25.2\,MB). These findings are consistent with the ``Jekyll and Hyde'' phenomenon documented by Aqeel et al.~\cite{aqeel2020landing}, where measurement choices about which pages to analyze can dramatically affect results.

\begin{table}[t]
\centering
\caption{Complete results for all 18 sites. Req = average total requests, API = average API calls, Size = average page size in KB, QS = quality score (0--100), RC = redundant call excess, MC\% = missing cache \%, TP\% = third-party request \%.}
\label{tab:full-results}
\small
\begin{tabular}{@{}l r r r r r r r@{}}
\toprule
\textbf{Site} & \textbf{Req} & \textbf{API} & \textbf{Size (KB)} & \textbf{QS} & \textbf{RC} & \textbf{MC\%} & \textbf{TP\%} \\
\midrule
News-1          & 681.8  & 87.8  & 9,314  & 56.8  & 13.0 & 27.4  & 95.8 \\
Utility-1       & 2,683.7& 243.7 & 11,275 & 59.9  & 73.5 & 39.0  & 96.7 \\
Commerce-3      & 433.2  & 62.3  & 4,934  & 60.8  & 26.2 & 58.8  & 85.0 \\
Travel-1        & 297.3  & 70.3  & 6,728  & 61.1  & 43.7 & 65.1  & 77.4 \\
Travel-2        & 248.3  & 50.5  & 4,358  & 61.8  & 15.2 & 31.9  & 84.3 \\
Entertainment-1 & 1,618.7& 131.3 & 8,089  & 63.6  & 25.7 & 36.9  & 98.7 \\
Commerce-1      & 426.2  & 36.3  & 5,937  & 65.3  & 10.8 & 81.9  & 31.2 \\
News-2          & 101.0  & 58.8  & 11,311 & 69.6  & 0.8  & 3.1   & 20.1 \\
News-3          & 971.5  & 56.5  & 25,817 & 70.9  & 8.2  & 37.4  & 60.8 \\
DevTool-2       & 349.3  & 25.5  & 2,431  & 78.7  & 3.8  & 24.4  & 79.0 \\
Commerce-2      & 231.7  & 32.7  & 4,683  & 81.4  & 0.0  & 14.0  & 88.1 \\
Reference-2     & 166.5  & 11.3  & 1,203  & 84.3  & 1.5  & 24.2  & 0.0  \\
DevTool-1       & 138.5  & 4.0   & 1,545  & 85.4  & 0.0  & 0.0   & 94.3 \\
Reference-1     & 59.5   & 0.5   & 1,365  & 92.2  & 0.0  & 0.0   & 51.8 \\
DevBlog-1       & 62.7   & 5.2   & 1,051  & 95.4  & 1.2  & 0.0   & 5.0  \\
Classifieds-1   & 68.5   & 3.0   & 1,612  & 97.5  & 0.0  & 0.0   & 0.0  \\
Government-1    & 13.5   & 0.0   & 182    & 100.0 & 0.0  & 0.0   & 0.0  \\
Forum-1         & 6.0    & 0.0   & 13     & 100.0 & 0.0  & 0.0   & 0.0  \\
\midrule
\textbf{Mean}   & 475.4  & 48.9  & 5,658  & 76.9  & 12.4 & 24.7  & 53.8 \\
\textbf{Median} & 240.0  & 34.5  & 4,521  & 74.8  & 2.6  & 24.3  & 69.1 \\
\bottomrule
\end{tabular}
\end{table}

\subsection{Anti-Pattern Prevalence}
\label{sec:antipatterns}

Figure~\ref{fig:heatmap} shows the anti-pattern heatmap across all 18 sites. The most pervasive anti-patterns are:

\begin{figure}[t]
    \centering
    \includegraphics[width=\textwidth]{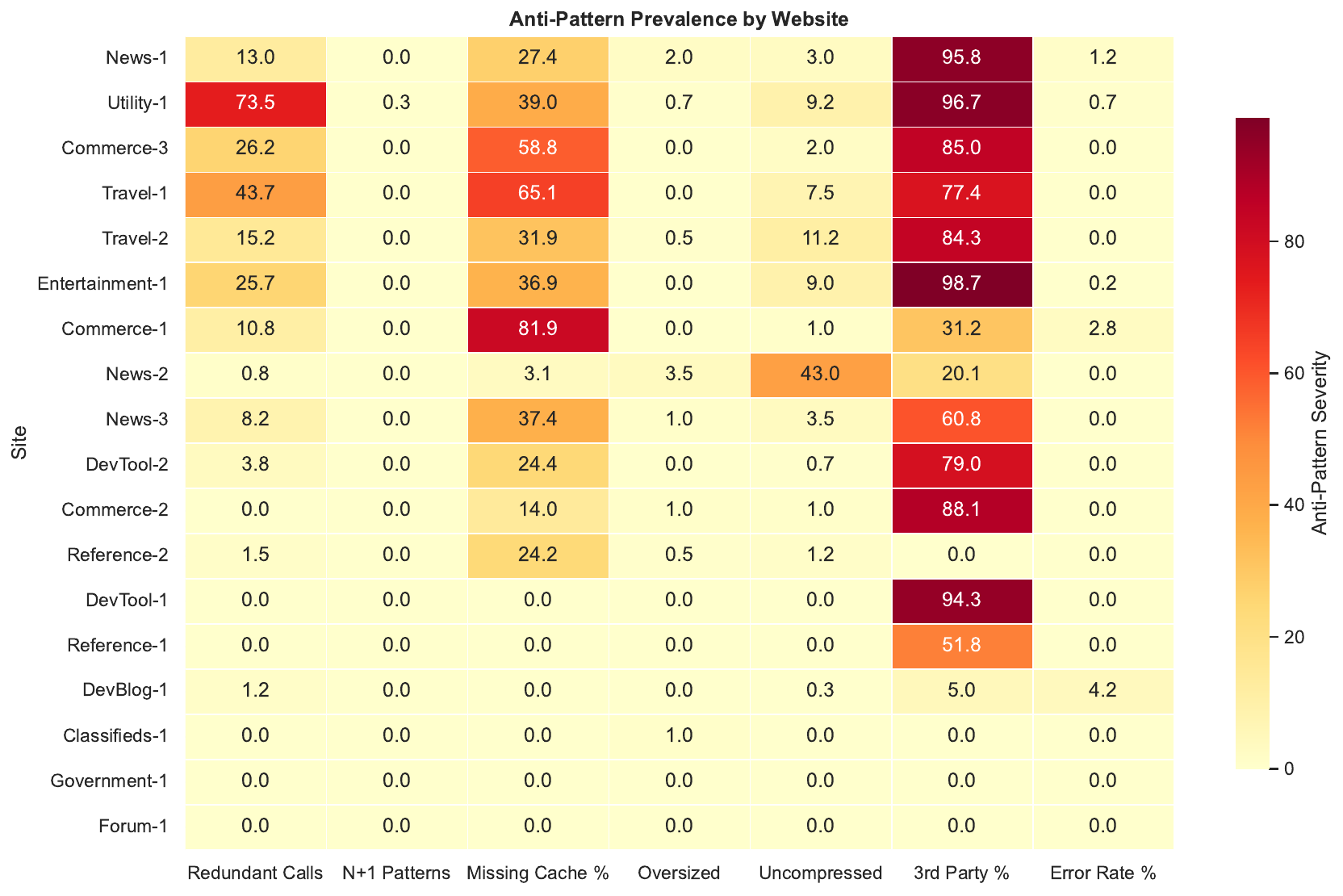}
    \caption{Anti-pattern prevalence heatmap. Darker cells indicate more severe issues. Each value represents the average count or percentage across all captures.}
    \label{fig:heatmap}
\end{figure}

\textbf{Third-party overhead} (mean 53.8\%): 13 of 18 sites (72\%) send more than 20\% of their requests to third-party domains. Entertainment-1 leads at 98.7\%, meaning nearly every request goes to a domain other than the site's own. Only 4 sites (Classifieds-1, Government-1, Forum-1, Reference-2) have zero third-party requests. This aligns with prior findings on the pervasiveness of third-party resources~\cite{nikiforakis2012you, kashaf2020analyzing}.

\textbf{Missing cache headers} (mean 24.7\%): 12 of 18 sites (67\%) have API responses lacking cache guidance. Commerce-1 is worst at 81.9\%, meaning over four-fifths of its API responses provide no \texttt{Cache-Control}, \texttt{ETag}, or \texttt{Last-Modified} header. This echoes the declining cache effectiveness trend reported by Ihm and Pai~\cite{ihm2011towards}.

\textbf{Redundant API calls} (mean 12.4 excess): 12 of 18 sites (67\%) make duplicate API calls. Utility-1 averages 73.5 excess redundant calls per page load, followed by Travel-1 (43.7) and Commerce-3 (26.2).

\textbf{Missing compression} (mean 5.1 uncompressed): News-2 stands out with 43 uncompressed API responses per page load, despite having relatively few total requests (101). This suggests large JSON payloads being served without gzip/brotli encoding, an area where compression proxies like Flywheel~\cite{agababov2015flywheel} have demonstrated 50\% savings.

\textbf{N+1 patterns} were rare (only Utility-1 showed 0.3 instances on average), and \textbf{error rates} were generally low (mean 0.5\%), with DevBlog-1 as an outlier at 4.2\%.

\subsection{Architecture vs. Quality}
\label{sec:architecture}

\begin{figure}[t]
    \centering
    \includegraphics[width=\textwidth]{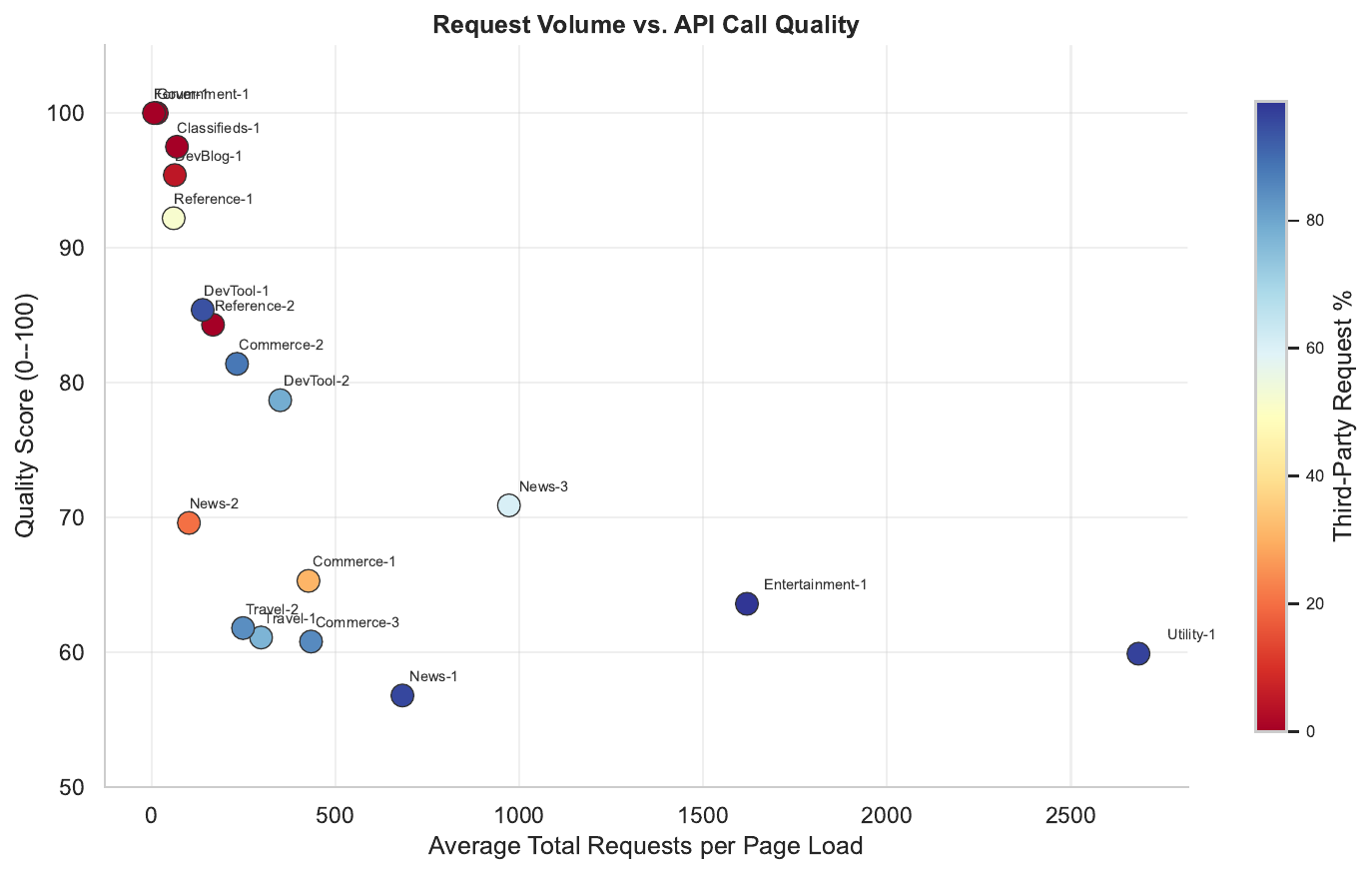}
    \caption{Request volume vs.\ quality score. Point color indicates third-party request percentage (blue = low, red = high). High-quality sites cluster in the bottom-left (few requests, low third-party).}
    \label{fig:scatter}
\end{figure}

Figure~\ref{fig:scatter} reveals a strong negative correlation between request volume and quality score. Sites making fewer than 100 requests consistently score above 90, while sites exceeding 500 requests all score below 75. This is consistent with the JavaScript performance overhead documented by Selakovic and Pradel~\cite{selakovic2016performance}.

The architectural pattern is clear: \textbf{server-rendered sites consistently outperform client-heavy SPAs}. The top 5 sites (Forum-1, Government-1, Classifieds-1, DevBlog-1, Reference-1) all use minimal or server-rendered architectures. The bottom 5 (News-1, Utility-1, Commerce-3, Travel-1, Travel-2) all employ heavy JavaScript frameworks with client-side hydration. This finding is consistent with systems research showing that static templating achieves 46--64\% faster loads by separating layout from dynamic API content~\cite{mardani2020fawkes}, and that JavaScript serial execution bottlenecks on modern pages are a dominant factor~\cite{mardani2021horcrux}. Iskandar et al.~\cite{iskandar2020comparison} observed similar trade-offs between client-side and server-side rendering approaches.

This does not mean server rendering is inherently superior, as the sites have different functional requirements. But it does suggest that JavaScript-heavy architectures \emph{amplify} the impact of poor API call practices: a redundant call in a 6-request page is negligible, but in a 2,684-request page it compounds with hundreds of others.

\subsection{Third-Party Ecosystem}
\label{sec:third-party}

Figure~\ref{fig:thirdparty} shows the third-party request breakdown by category. Among sites with third-party requests, the dominant categories are:

\begin{figure}[t]
    \centering
    \includegraphics[width=\textwidth]{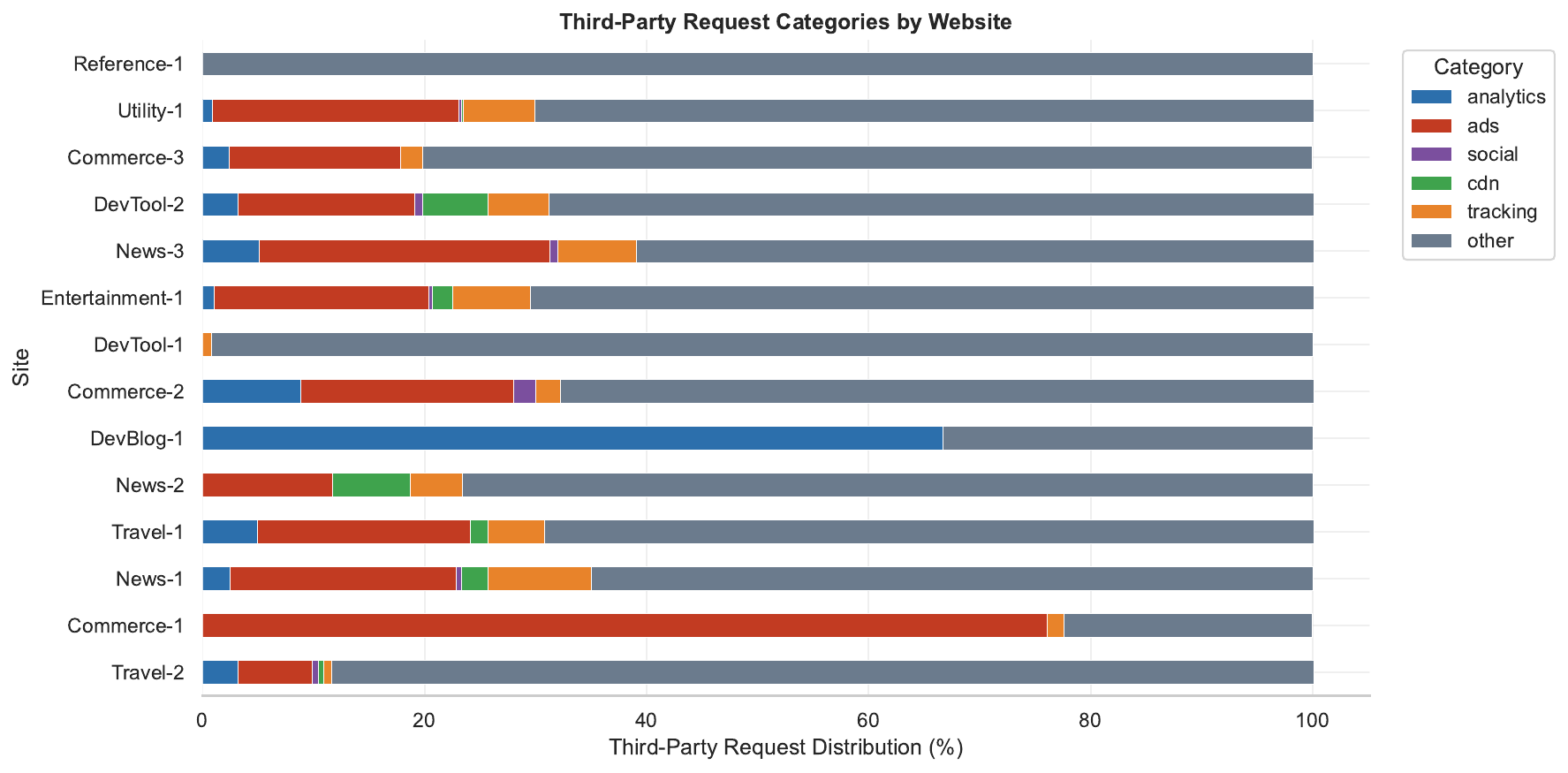}
    \caption{Third-party request distribution by category. Analytics and ads dominate the third-party ecosystem for most commercial sites.}
    \label{fig:thirdparty}
\end{figure}

\begin{itemize}[leftmargin=2em]
    \item \textbf{Analytics/tracking}: Present on 15 of 18 sites (83\%), often constituting the largest share. This prevalence is consistent with the census findings of Englehardt and Narayanan~\cite{englehardt2016census}.
    \item \textbf{Advertising}: Dominant on ad-supported sites (Utility-1, News-1, Entertainment-1), with complex chains of ad auctions and pixel fires reflecting the diffusion patterns documented by Bashir and Wilson~\cite{bashir2018diffusion}.
    \item \textbf{CDN}: Sites like DevTool-1 (94.3\% third-party) route nearly all content through CDN domains that technically register as ``third-party'' under our domain-matching heuristic.
    \item \textbf{Social}: Minimal presence, as most sites have moved away from embedded social widgets.
\end{itemize}

\subsection{Per-Capture Variability}
\label{sec:variability}

Across 3 independent runs, request counts show moderate variability. For example, DevTool-2's question page varied from 421 to 1,123 requests across 3 runs (likely due to ad network auction timing), while Forum-1 was perfectly stable at 6 requests per run. Sites with heavy advertising (Utility-1, News-1, Entertainment-1) showed the highest run-to-run variance, confirming that ad networks introduce significant non-determinism into page loads, a phenomenon characterized by Goel et al.~\cite{goel2024sprinter} as JavaScript-induced non-determinism in web crawls. This challenge has been documented in web measurement reproducibility studies~\cite{demir2022reproducibility, demir2023similarity}, where even identical measurement configurations produce divergent results across runs.

\subsection{Best vs. Worst Comparison}
\label{sec:comparison}

\begin{figure}[t]
    \centering
    \includegraphics[width=\textwidth]{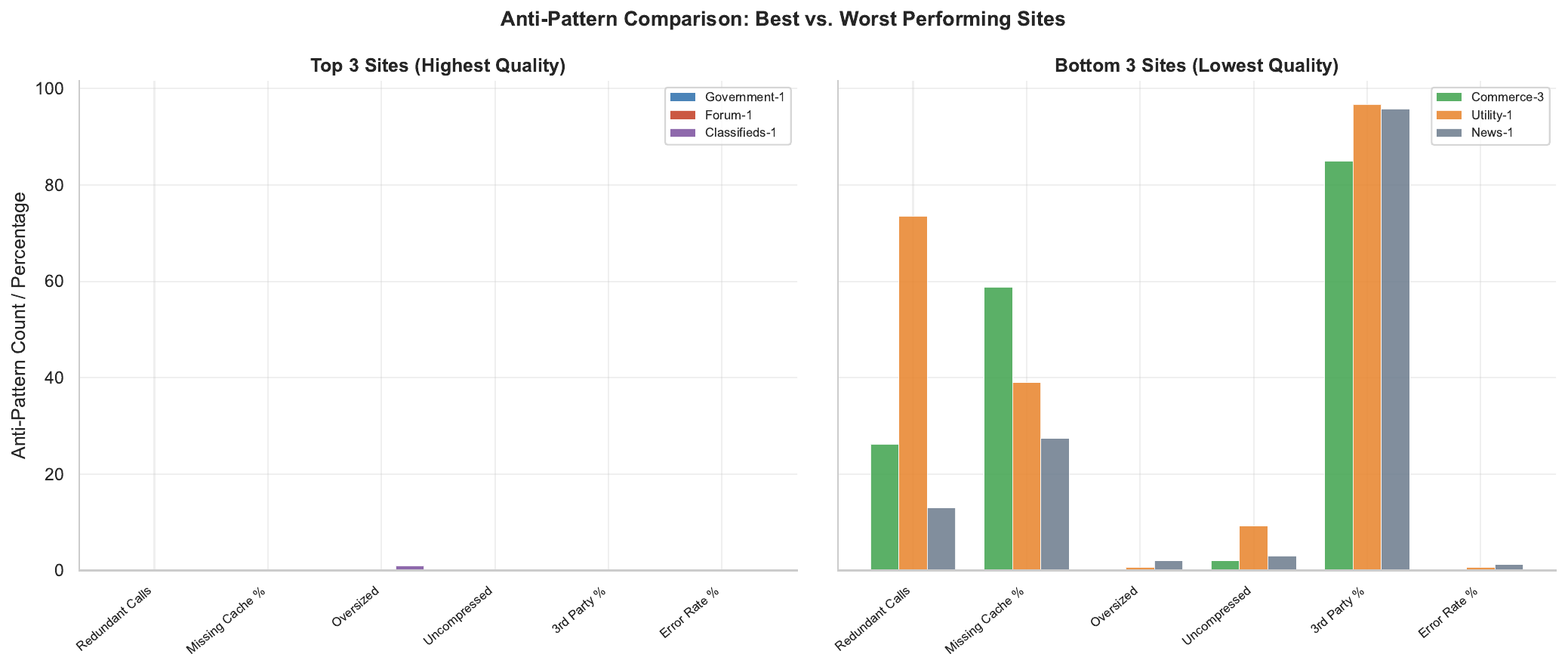}
    \caption{Anti-pattern comparison between the top 3 (highest quality) and bottom 3 (lowest quality) sites. Best-performing sites show uniformly near-zero anti-patterns.}
    \label{fig:bestworst}
\end{figure}

Figure~\ref{fig:bestworst} contrasts the anti-pattern profiles of the top 3 and bottom 3 sites. The top performers (Forum-1, Government-1, Classifieds-1) have uniformly zero anti-patterns across all dimensions. The bottom performers (News-1, Utility-1, Commerce-3) show elevated issues across multiple dimensions simultaneously. It is not a single anti-pattern dragging scores down, but rather a combination of redundant calls, missing cache headers, and third-party overhead.

\section{Discussion}
\label{sec:discussion}

\subsection{Answers to Research Questions}

\textbf{RQ1: How good are the HTTP API call patterns of popular production websites?} Quality scores range from 56.8 to 100.0 (mean 76.9, median 74.8). Minimal server-rendered sites consistently achieve scores above 90, while JavaScript-heavy commercial sites score below 75. The most pervasive anti-patterns are redundant API calls and missing cache headers, each affecting 67\% of sites, followed by third-party overhead exceeding 20\% on 72\% of sites.

\textbf{RQ2: What are the quality and security implications of the anti-patterns observed?} The detected anti-patterns carry direct security implications: excessive third-party dependencies (up to 98.7\%) expand the supply-chain attack surface; missing cache headers (up to 81.9\%) enable cache poisoning by intermediary proxies; error responses risk leaking server internals; and oversized payloads may expose data fields not rendered by the client. From a quality perspective, we find a strong correlation between architectural complexity and lower scores, indicating a measurable ``third-party tax'' and ``JavaScript complexity tax'' on API quality.

\subsection{Key Findings}

\textbf{Finding 1: The web has a wide quality spectrum.} Quality scores range from 56.8 to 100.0. This ${\sim}$43-point gap exists among the world's most popular websites, suggesting that even well-resourced engineering teams ship suboptimal API patterns.

\textbf{Finding 2: Redundant calls and missing cache headers are endemic.} These two anti-patterns affect the majority of sites. Redundant calls are easy to introduce (for example, a React component re-mounting triggers duplicate fetches), and missing cache headers suggest that backend teams may not be coordinating with frontend teams on caching strategy. This finding resonates with the observation by Vesuna et al.\ that caching provides less benefit than expected due to dynamic content~\cite{vesuna2016caching}.

\textbf{Finding 3: Third-party overhead is pervasive but varies dramatically.} From 0\% (Classifieds-1, Government-1, Forum-1, Reference-2) to 98.7\% (Entertainment-1). Commercial sites pay a heavy ``third-party tax'' through analytics, advertising, and tracking integrations. The resilience implications of these dependencies are significant, as Kashaf et al.~\cite{kashaf2020analyzing} demonstrated following the Mirai-Dyn incident.

\textbf{Finding 4: Minimal architectures achieve higher quality scores, though they face different constraints.} The correlation between fewer requests and higher scores is strong. However, we acknowledge that minimalist sites do not need to run real-time ad auctions, personalization engines, or A/B testing frameworks that commercial sites require. The trade-off between functionality and API quality deserves further study.

\subsection{Practical Implications}

For \textbf{web developers}: The most impactful improvements are (1) deduplicating API calls at the client level (e.g., using SWR, React Query, or Apollo Client caching), (2) adding cache headers to API responses~\cite{vesuna2016caching}, and (3) auditing third-party scripts for necessity~\cite{nikiforakis2012you}.

For \textbf{engineering managers}: Our scoring framework can serve as a continuous integration check. By capturing HAR files in CI, running the anti-pattern detectors, and alerting when scores drop below a threshold, teams can catch quality regressions before deployment. This approach complements existing web testing practices~\cite{mughal2025autonomous, li2024survey, kertusha2025survey}, extending quality assurance from functional correctness to network efficiency. Combining UI-layer testing~\cite{mughal2025autonomous} with API-layer quality auditing could provide end-to-end web application quality monitoring.

For \textbf{researchers}: Our open dataset enables further studies on API quality trends over time, cross-country variation, or correlations with user experience metrics (Core Web Vitals~\cite{webvitals2020}, Time to Interactive). The framework is extensible: new anti-pattern detectors can be added to the scoring rubric.

\subsection{Security Implications}
\label{sec:security}

Several of the quality anti-patterns we detect carry direct security implications, linking software quality testing to security assessment at the network layer.

\textbf{Third-party supply-chain risk.} Sites with high third-party request percentages (e.g., Entertainment-1 at 98.7\%) depend on external domains they do not control. If any third-party analytics provider, ad network, or CDN is compromised, it can inject malicious JavaScript into every page that includes it. The Mirai-Dyn incident~\cite{kashaf2020analyzing} demonstrated that concentrated third-party dependencies create systemic fragility, and Ikram et al.~\cite{ikram2019trust} showed that approximately 50\% of first-party websites render content loaded through deep transitive trust chains, further amplifying supply-chain exposure. Our third-party overhead scores quantify this attack surface: a site with 0\% third-party requests has zero supply-chain exposure through its API layer, while a site at 98.7\% has nearly complete exposure.

\textbf{Cache poisoning via missing headers.} API responses without \texttt{Cache-Control} headers are susceptible to cache poisoning by intermediary proxies. When a response lacks explicit caching directives, proxy servers and CDN edge nodes may cache it with default policies, potentially serving stale or manipulated data to subsequent users. Commerce-1, with 81.9\% of API responses missing cache headers, is particularly vulnerable. Proper \texttt{Cache-Control: no-store} directives on sensitive endpoints and validation headers (\texttt{ETag}, \texttt{Last-Modified}) on cacheable responses would mitigate this risk.

\textbf{Information leakage through error responses.} HTTP 4xx and 5xx responses can inadvertently reveal server implementation details such as framework versions, internal file paths, stack traces, and database error messages. While error rates in our corpus are generally low (mean 0.5\%), even a small number of verbose error responses can provide reconnaissance information to attackers. DevBlog-1's 4.2\% error rate warrants investigation to ensure that error responses are properly sanitized.

\textbf{Over-fetching and data exposure.} Oversized API payloads (detected by D5) may transfer data fields that the client does not render but that an attacker could exploit. This violates the principle of least privilege at the data transfer level: API responses should contain only the fields required by the consuming client. GraphQL adoption~\cite{brito2020rest, seabra2019rest} can mitigate this by allowing clients to specify exactly which fields they need.

These findings suggest that network-layer quality testing, as implemented by our framework, can serve as a lightweight security audit that complements traditional penetration testing and vulnerability scanning.

\section{Threats to Validity}
\label{sec:threats}

\textbf{External validity.} Our corpus of 18 sites, while diverse, is not a random sample of the web. Results may not generalize to all websites, particularly those in non-English markets or behind authentication walls. Xavier~\cite{xavier2024web} found that web traffic is highly concentrated (50\% of traffic goes to ~3,000 sites), so our sample of popular sites captures behavior relevant to a large share of actual web usage.

\textbf{Measurement validity.} Automated browsing may trigger different server behavior than human visitors (e.g., fewer personalized ads, different A/B test variants). We mitigated this with realistic user-agent strings and scrolling behavior~\cite{chiapponi2023detecting}, but cannot fully replicate human interaction. Jueckstock et al.~\cite{jueckstock2021crawl} showed that crawling setup choices significantly affect measurement results, and Goel et al.~\cite{goel2022jawa} demonstrated that JavaScript-heavy pages are inherently difficult to capture faithfully. Demir et al.~\cite{demir2022reproducibility} note this as a common limitation of web measurement studies.

\textbf{Heuristic accuracy.} Our API call identification relies on Content-Type and URL pattern heuristics. Some API calls using non-standard formats may be missed, and some non-API requests (e.g., JSON configuration files) may be false positives. Similar trade-offs are inherent in automated anti-pattern detection~\cite{palma2014detection, sharma2018survey}. Additionally, specific detectors carry inherent limitations: D3 (Sequential Waterfalls) cannot distinguish legitimately dependent call chains (e.g., an authentication token request followed by an authorized API call) from parallelizable requests, since HAR files do not encode dependency relationships. D4 (Missing Cache Headers) cannot differentiate between responses that are intentionally uncacheable (e.g., personalized pricing, real-time data) and those where cache headers were simply omitted. In both cases, domain-specific knowledge would be needed to eliminate false positives.

\textbf{Third-party classification.} We use simple domain substring matching to classify first-party vs.\ third-party requests. This may misclassify some corporate CDN domains as third-party when they use separate domain names from the main site. A more sophisticated approach using the Public Suffix List, corporate domain mappings, or graph-based classification such as AdGraph~\cite{iqbal2020adgraph} (which achieves 95\% accuracy) would improve precision. Furthermore, D7 penalizes all third-party requests equally, without distinguishing between functionally necessary dependencies (e.g., payment processors, authentication providers) and discretionary ones (e.g., ad networks, social widgets). A weighted classification that accounts for functional necessity would yield more nuanced scores.

\textbf{Scoring subjectivity.} The dimension weights and penalty curves were chosen based on our assessment of practical impact. Different weights would produce different rankings. We provide per-dimension scores to enable alternative weighting schemes. Notably, D2 (N+1 Patterns) exhibited near-zero prevalence across all 18 sites (mean 0.0, max 0.3), meaning its 10\% weight contributes no discriminatory power; this pattern, common in backend code, is likely masked by server-side aggregation layers and is fundamentally difficult to detect at the network level. Similarly, D5 (Oversized Payloads) had low prevalence (mean 0.6, max 3.5). Together, D2 and D5 account for 25\% of the composite weight yet contribute minimal differentiation, effectively inflating all scores. The penalty thresholds (100\,KB for oversized payloads, 1\,KB for compression) follow common industry practice but are not empirically derived for all application types.

\textbf{Temporal validity.} Websites change frequently. Our captures represent a snapshot from February 2026. Request counts and anti-pattern prevalence may differ at other times.

\section{Reproducibility}
\label{sec:reproducibility}

Following reproducibility best practices~\cite{demir2022reproducibility, wohlin2012experimentation}, all code and anonymized data are publicly available at:

\begin{center}
\url{https://github.com/amughalbscs16/network-layer-quality-testing}
\end{center}

The repository contains:

\begin{itemize}[leftmargin=2em]
    \item \texttt{capture.js}: Node.js/Playwright script for HAR capture (run with \texttt{node capture.js --accessible-only}).
    \item \texttt{analyze.py}: Python analysis pipeline with 8 anti-pattern detectors and composite scoring (run with \texttt{python analyze.py --anonymize}).
    \item \texttt{validate.py}: Independent data validation with 8 automated checks (run with \texttt{python validate.py}).
    \item \texttt{data/sites\_anonymized.json}: Website corpus with anonymized metadata (URLs redacted).
    \item \texttt{results/}: Anonymized CSV tables (quality scores, anti-pattern counts, summary statistics) and 18 per-site JSON reports with full dimension breakdowns.
\end{itemize}

\textbf{Note on raw data}: Raw HAR files are not included in the public repository as they contain full HTTP request/response headers with identifiable domain names. They are available upon request for verified academic research purposes.

\textbf{Dependencies}: Node.js 18+, Playwright 1.49+, Python 3.10+, pandas, matplotlib, seaborn.

\section{Conclusion and Future Work}
\label{sec:conclusion}

We presented a systematic, automated software testing approach for assessing HTTP API call quality across 18 popular production websites. Our three-phase pipeline (capture, analyze, score) reveals significant quality variation, with scores ranging from 56.8 to 100.0. The most pervasive issues are redundant API calls and missing cache headers, both of which are straightforward to fix with modern client-side caching libraries and proper server configuration. Beyond performance, we demonstrated that several quality anti-patterns carry direct security implications, including supply-chain exposure through third-party dependencies and cache poisoning risks from missing cache headers.

The strong correlation between architectural complexity and lower quality scores suggests that the ``third-party tax'' and ``JavaScript complexity tax'' are real and measurable. As web applications continue to grow in complexity, systematic quality and security auditing of the kind we demonstrate here will become increasingly important.

\textbf{Future work.} We plan to extend this study in four directions: (1) feeding captured HAR data to multiple Large Language Models and comparing their ability to diagnose API anti-patterns against our heuristic baseline; (2) repeating the audit longitudinally to track quality trends over time; (3) integrating our network-layer API quality analysis with UI-layer testing approaches~\cite{mughal2025autonomous} to create a comprehensive, multi-layer web application quality framework; and (4) expanding the security analysis dimension with dedicated detectors for OWASP API Security Top 10 risks, including broken authentication tokens in URLs, excessive data exposure in API responses, and insecure transport configurations.

\section*{About the Authors}

\textbf{Ali Hassaan Mughal} is an independent researcher currently pursuing an Applied MBA in Data Analytics at Texas Wesleyan University, USA. He earned his M.Sc.\ in Computer Science from Kansas State University. Prior to his current research endeavors, he served as a Senior Software Developer and Team Lead at Xpressdocs and as a Software Developer~II at Paycom. His research interests include automated software testing, web application quality assurance, and applied machine learning.

\textbf{Muhammad Bilal} is an independent researcher pursuing an M.Sc.\ in Management at the Technical University of Munich, Germany. He has professional experience as both a Business Analyst and a Software Engineer. His research interests focus on the impact of technology on business performance, product quality analytics, and the automation of industrial pipelines and processes.

\textbf{Noor Fatima} is an undergraduate student pursuing a B.E.\ in Computer Engineering at the National University of Sciences and Technology (NUST), Pakistan. She has a strong interest in software development and hardware systems, with a growing focus on emerging technologies in computing.

\bibliographystyle{IEEEtran}
\bibliography{refs}
\end{document}